\documentclass[aps,pre,showpacs,twocolumn]{revtex4-1}

\usepackage{graphicx}
\usepackage{latexsym}

\begin{document}

\title{Dynamics of periodic node states on a model of static networks with
repeated-averaging rules}

\author{Suhan Ree}
\email{suhan@physics.utexas.edu}
\affiliation{Center for Complex Quantum Systems and Department of Physics, University of Texas at Austin, Austin, TX, 78712, USA}
\affiliation{School of Liberal Arts and Science, Kongju National University, Yesan-Up, Yesan-Gun, Chungnam, 340-702, South Korea}

\date{\today}

\begin{abstract}
We introduce a simple model of static networks, where
nodes are located on a ring structure, and two accompanying 
dynamic rules of repeated averaging on periodic node states.
We assume nodes can interact with neighbors, and will add long-range
links randomly.
The number of long-range links, $E$, controls structures of these networks, and
we show that there exist many types of fixed points, when $E$ is varied.
When $E$ is low, fixed points are mostly diverse states, in which 
node states are diversely populated;
on the other hand, when $E$ is high, fixed points tend to be dominated 
by converged states, in which node states converge to one value.
Numerically, we observe properties of fixed points for various $E$'s, and also 
estimate points of the transition from diverse states to converged states
for four different cases.
This kind of simple network models will help us understand how diversities that 
we encounter in many systems of complex networks are sustained, even when
mechanisms of averaging are at work,
and when they break down if more long-range connections are added.
\end{abstract}



\maketitle

\section{Introduction}
Complex networks have been a popular subject of statistical physics for more than a decade,  
because of their versatility and rich dynamic behavior\cite{Newman:2003,Dorogovtsev:2008}. 
In addition, they might be the only way to explain some of the universal 
features of various systems that can be represented by networks.
Even though structures of static networks can be a research subject of its own,
studies on dynamic networks are also flourishing.
Three approaches can used on models of network dynamics, 
depending on the characteristics of systems in question:
(i) node dynamics, focusing on dynamics of node states, while link structures are static,
(ii) link dynamics, focusing on dynamics of links and their structures, while 
nodes are stateless, or have static states, and
(iii) coevolving dynamics, where both links and nodes evolve, possibly influencing each other.
Note that when some entities are assumed to be {\em static}, it means that
the characteristic times for their changes are considerably longer than
those of {\em dynamic} entities.
One field that has attracted many physicists recently as an application of 
dynamic complex networks 
is a study of social systems\cite{Galam:1982,Castellano:2009}, where nodes are humans or 
groups
of humans, and their relationships define links, forming social networks.
Social entities influence each other, mostly in attracting ways, but 
one of the fundamental characteristics of our societies is the sustained 
diversities. 
How are they being sustained when social entities have tendencies 
to assimilate with each other?
How and when do these diversities break down?
These are the questions we try to answer using complex networks
on problems of
opinion dynamics\cite{Chatterjee:1977,Hegselmann:2002,Weisbuch:2002,Amblard:2004,SznajdWeron:2000,Mas:2010,Ree:2011},
cultural dynamics\cite{Schelling:1969,Axelrod:1997,Laguna:2003} and so on.

Here we introduce a model of static networks and investigate 
node dynamics on these networks using numerical simulations mostly,
while we have already studied the same dynamic rule 
on dynamic networks\cite{Ree:2011,Ree:2011a}.
This model has nodes with periodic states
with repeated-averaging rules\cite{Feller,Chatterjee:1977}, and
there are two types of links: short-range and long-range links.
Long-range interactions are rapidly increasing in our societies, as we are recently 
seeing with ever-advancing technologies, and we will use the number of long-
range links, $E$, as the control 
parameter of the model.
In our model, $E$ changes the network structure,
and has an important role in determining what types of fixed points are 
reached from random initial conditions.
When there are not many long-ragne links, diverse states prevail, but
diversities will disappear eventually and more converged states will appear.  
As we vary $E$, we will find several types of fixed points, 
and estimate the likelihood of spontaneous 
emergence for given types of fixed points, using Monte-Carlo simulations.
Network models with periodic nodes can be applied not only 
to social systems, but also
to other domains as in synchronization problems in oscillator networks\cite{Arenas:2008}.
For example, there have been many works on dynamics of
Kuramoto model\cite{Kuramoto:1984,Acebron:2005} on complex 
networks\cite{Zheng:1998,Hong:2002,Rogge:2004,GomezGardenes:2007,Verwoerd:2009,DiazGuilera:2009,Kalloniatis:2010}, 
and one dynamic rule in our model is similar to that of Kuramoto model.

The plan of this paper is as follows.
In Sec.\ II, the network model and two dynamic rules are introduced, 
and dynamics of networks with a small number of nodes is examined.
In Sec.\ III, numerical results are presented. We find different types of fixed points using Monte-Carlo 
simulations, and observe their properties as $E$ varies. 
Finally, in Sec.\ IV, we conclude with discussions.
 
\section{Model}
\begin{figure}
\begin{center}
\includegraphics[scale=0.50]{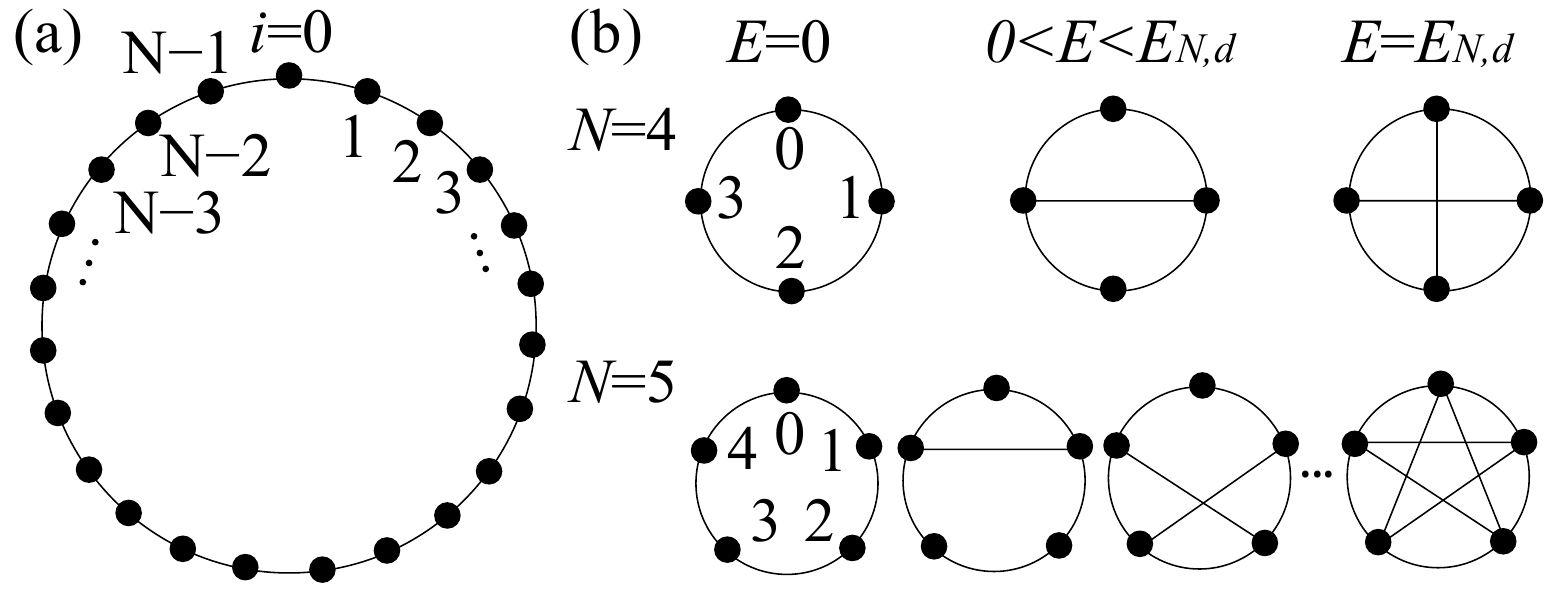}
\caption{\label{model}
	(a) $N$ nodes in an 1D periodic lattice with only nearest-neighbor 
	connections ($d=1$ here). 
	(b) Examples of $N=4$ and 5 when $d=1$. 
	We can add $E$ long-range links,
	 where $E_{N,d}$ is the maximum number of long-range connections
	 for given $N$ and $d$.}
\end{center}
\end{figure}

We describe a network model with simple examples in Fig.\ \ref{model}.
In this model, the network has $N$ nodes, and 
they are placed in a ring structure, a 1D periodic lattice.
Each node is connected to neighbors within distance $d$, assuming $N\gg d$;
then, there are $dN$ links, called {\em short-range} links.
Now we add $E$ distinct links randomly, not between already-connected 
neighbors, 
and these links are {\em long-range} links (or {\em shortcuts}).
The total number of links becomes $dN+E$, and the average connectivity 
$k$ is $2(d+E/N)$.
If we introduce the normalized parameter $e\equiv E/E_{N,d}$, where 
$E_{N,d}$ is $N(N-1-2d)/2$, the maximum number of distinct long-range links,
this becomes a normalized control parameter of the network ($0\le e\le1$)\footnote{
We can also use the random network of 
Erd{\"o}s and R{\'e}nyi
on top of the regular network with only short-range links. 
The wiring probability $p$ ($0\le p\le1$)
becomes the control parameter, and results will not be much different
when $d$ is small.}.
This model can be viewed as a modified version of the small-world network 
by Newman and Watts\cite{Newman:1999}.
When $d=0$, networks will have the same structural properties as random
networks while $E$ is the total number of links.
When $e=0$ and $d>1$, the network is a regular network, and,
as $e$ increases, the network undergoes several structural phases including small-world networks for low $e$.

\begin{figure}
\begin{center}
\includegraphics[scale=0.5]{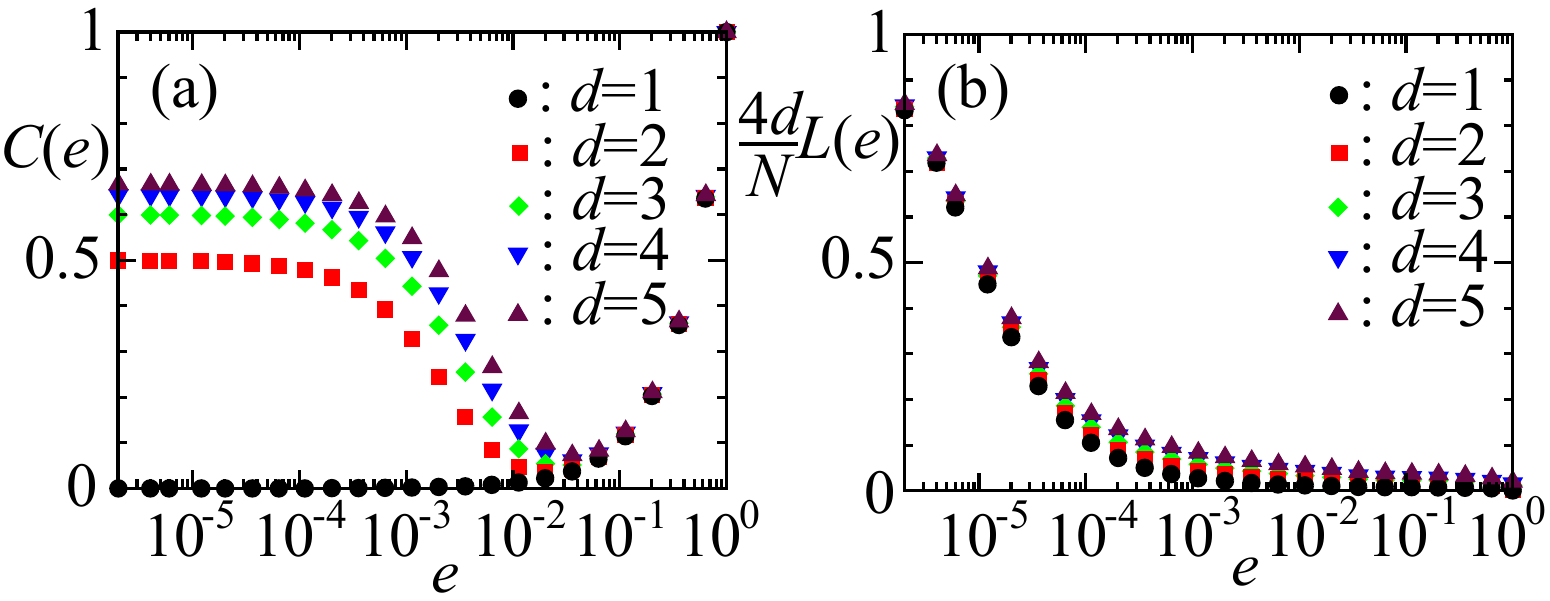}
\caption{\label{smallworld}
	(Color online)
	For $N=1000$, we observe (a) the clustering coefficient, $C(e)$, and (b)
	the average shortest path length, $L(e)$, versus $e$ for
	$1\le d\le5$ (averaged over 1000 random network configurations).}
\end{center}
\end{figure}

In Fig.\ \ref{smallworld},
we observe the clustering coefficient $C(e)$ and the average
shortest path length $L(e)$ as we vary $e$, for the cases for $N=1000$ and 
$1\le d \le 5$.
When $e=0$ and $d\ge1$, we can easily find $C(0)$ and $L(0)$ as below,
\begin{eqnarray}
C(0)&=&\frac{3(d-1)}{2(2d-1)},\\
L(0)&=&\left(\frac{q+1}{N-1}\right)\left(\frac{N-\delta}{2}+r\right)
		\simeq\frac{N}{4d},
\end{eqnarray}
where  $q$ and $r$ are the quotient and the remainder when dividing $(N-2+\delta)/2$ 
by $d$, and $\delta$ is defined as
\begin{equation}
\delta\equiv\left\{\begin{array}{l}
			0 \mbox{\ ($N$: even),}\\
			1 \mbox{\ ($N$: odd).}
		\end{array} \right.
\end{equation}
For $d>1$, $C(e)$ does not change much until $e$ reaches 
$2\times10^{-4}$ ($E\simeq100$), while $L(e)$ is reduced to about $L(0)/10$.
As $e$ increases further, $C(e)$ decreases sharply to almost 0 and starts
to increase at $e\simeq2\times10^{-2}$ ($E\simeq10^4$), and finally reaches the fully-connected network at $e=1$. 
On the other hand, $L(e)$ monotonically decreases until it reaches 1 at $e=1$.

Now we introduce a state variable $\phi_i$ to each node $i$, and assume
$\phi_i$ is periodic with the range of $0\le\phi_i<1$.
Then the state space of all $N$ nodes, which can be represented 
by an $n$-tuple, $(\phi_0,\phi_1,\phi_2,\ldots,\phi_{N-1})$,
becomes an $N$-dimensional torus ($N$-torus).
Now we impose a synchronous and deterministic dynamic rule with discrete 
time $t$ ($t=0,1,2,\ldots$), and the rule can be represented by a map
for each node $i$ at each time step as below,
\begin{equation}
	\phi_i \leftarrow \phi_i+\frac{\sigma}{k_i+1}\sum_{j=0}^{N-1} K_{ij}
	\Delta(\phi_j-\phi_i)\ \mbox{(mod 1)}, 
	\label{phi_change}
\end{equation}
where $\sigma$ is a coupling constant ($0<\sigma\le1$, but we will assume
$\sigma=1$ for simplicity here),
$K_{ij}$ is an element of the adjacency matrix (1 when there 
exists a link between $i$ and $j$, and 0 otherwise), 
$k_i$ is the degree of node $i$,
and $\Delta(\phi_j-\phi_i)$ represents
the difference between $\phi_j$ and $\phi_i$.
Then, the next value is the 
{\em average} of all states of neighbors of $i$ and $i$ itself when $\sigma=1$.
(If $\sigma<1$, $\phi_i$ will move closer to the average, and, if $\sigma>1$, which is not allowed here, 
the next value of $\phi_i$ will overshoot the average.)

If the state variable is not periodic, we can simply use $\Delta(x)=x$.
But the state variable is periodic here,
and the definition of averaging can become ambiguous.
For example, if we have two values, 0.1 and 0.9,
there are two middle points, 0 and 0.5, and which is the {\em average}?
The answer is both can be right.
If we calculate the average from the point of view at, let's say, 0.45, the average is
0.5 because distances to two points from 0.45 are -0.35 and 0.45, respectively.
But from the point of view at 0.1, the average becomes 0, because distances to two points
are 0 and -0.2.
Then, an average in view of one value can be different from those in view of other values.
This means that we need an anchor value to find an average, and that averages are relative.
In Eq.\ (\ref{phi_change}), each node $i$ is using its state value $\phi_i$ as the anchor,
and is finding its own average at each time step.
But there still exists an ambiguity when distances between two values are exactly 0.5; 
for example, if the anchor value is 0 and a value is at 0.5, the distance can be either 0.5 or -0.5.
But, if we set $\Delta(x)=0$ at $x=\pm0.5$, this ambiguity can be avoided. 
In addition, $\Delta(x)$ should be periodic [$\Delta(x\pm1)=\Delta(x)$ for $-1<x<1$], 
because $x$ is a periodic variable.

\begin{figure}
\begin{center}
\includegraphics[scale=0.35]{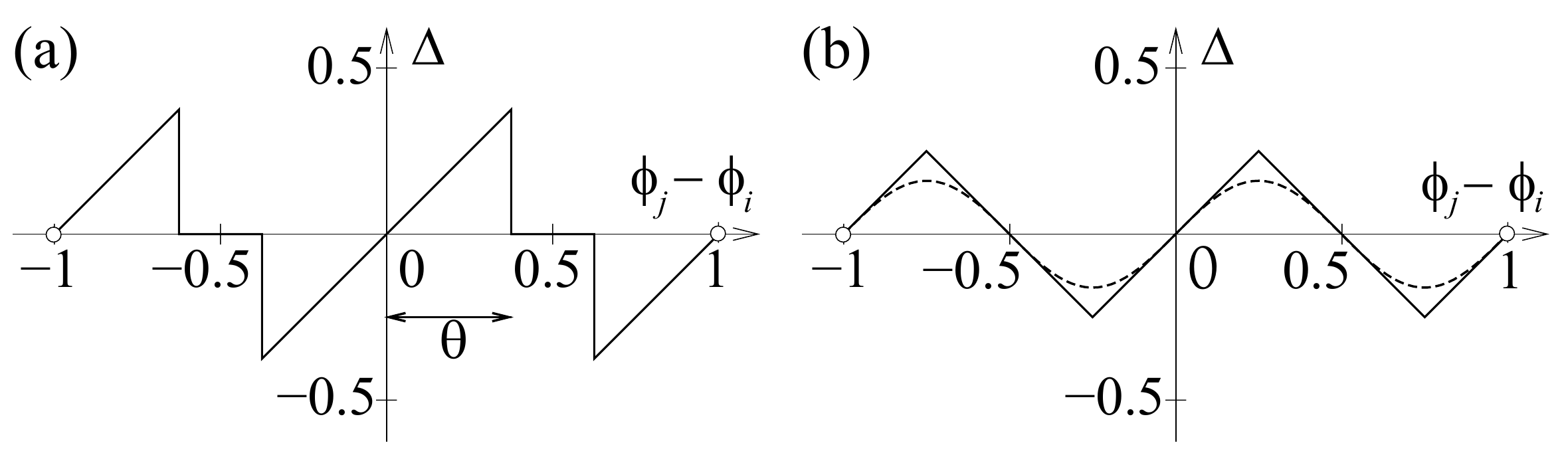}
\caption{\label{delta}
	The function $\Delta(\phi_j-\phi_i)$ representing the difference 
	between $\phi_j$ and $\phi_i$. 
	(a) Rule A with a threshold $\theta$ in Eq.\ (\ref{periodic1}).
	(b) Rule B in Eq.\ (\ref{periodic2}). 
	The dashed line is $\sin[2\pi(\phi_j-\phi_i)]/(2\pi)$, 
	as in the Kuramoto model. 
}
\end{center}
\end{figure}

One way to define $\Delta(x)$ is as below [see Fig.\ \ref{delta}(a)],
\begin{equation}
	\Delta_A(x)=\left\{
		\begin{array}{ll}
			x+1 & (\mbox{if}\ x<-1+\theta),\\
			0 & (\mbox{if}\ -1+\theta\le x\le -\theta),\\
			x & (\mbox{if}\ -\theta<x <\theta),\\
			0 & (\mbox{if}\ \theta\le x\le 1-\theta),\\
			x-1    & (\mbox{if}\ 1-\theta<x), 
		\end{array}\right.
	\label{periodic1}
\end{equation}
where $\theta$ is the threshold for interaction ($0<\theta\le0.5$),
and we will call this rule as Rule A.
Thresholds have been used in models of {\em bounded confidence}\cite{Deffuant:2000,Hegselmann:2002,Weisbuch:2002,Amblard:2004} in 
opinion dynamics, and
can influence the dynamics greatly. 
In this work, we set $\theta$ as 0.5, the maximum value, for simplicity.
Now we can find the next value of $\phi_i$ unambiguously. 
What happens if we repeatedly find averages using Eq.\ (\ref{phi_change}) for a given initial condition
for $N$ nodes?
Values will eventually converge to a fixed point in the state space, where
the average among neighbors for each node is the same as the state value of itself.
This rule has been used in previous works by the author\cite{Ree:2011,Ree:2011a}. 
Note that, due to the discontinuities in $\Delta_A(x)$,
a slight change of 
the state for one node can induce abrupt changes of other nodes at the next time step.

We can avoid these discontinuities, and introduce
another form of $\Delta(x)$ by modifying Eq.\ (\ref{periodic1})
as shown in the next equation [see Fig.\ \ref{delta}(b)],
\begin{equation}
	\Delta_B(x)=\left\{
		\begin{array}{ll}
			x+1 & (\mbox{if}\ x\le-0.75),\\
			-x-0.5 & (\mbox{if}\ -0.75<x\le-0.25),\\
			x & (\mbox{if}\ -0.25<x \le0.25),\\
			-x+0.5 & (\mbox{if}\ 0.25<x\le0.75),\\
			x-1    & (\mbox{if}\ 0.75<x), 
		\end{array}\right.
	\label{periodic2}
\end{equation}
and we will call this rule as Rule B.
This rule resembles the interaction term in 
the Kuramoto model\cite{Kuramoto:1984}, and we can use another form of $\Delta(x)$ as below,
\begin{equation}
	\Delta_K(x)=\frac{1}{2\pi} \sin(2\pi x),
	\label{kuramoto}
\end{equation}
where results will be similar to those from Rule B (will not be shown here). 
Under both rules, A and B, each node updates its state with the 
{\em averaged} value especially when $|\phi_j-\phi_i|<0.25$, 
and we can call both of them as the rules of repeated averaging of periodic variables.
One difference is that $|\Delta_B|$ decreases as $|x|$ increases from 0.25 to 0.5 unlike Rule A,  
making the meaning of averaging a little different.
This can be interpreted as another representation of the concept of bounded confidence, because the
influence between nodes starts to decrease when the difference is greater than 0.25.
Later we will see that this fact can also play an important role for 
stability of some fixed points.

We know that the {\em converged} states (or {\em consensus}
if we interpret $\phi_i$ as an opinion of node $i$, or {\em synchronized} states if 
we interpret $\phi_i$ as a phase of one of identical oscillators), 
where all nodes have the same 
value, are always stable fixed points under both rules.
But there are other kinds of fixed points, and one type of fixed points
can be represented by 
\begin{equation}
	\phi_i=\phi_0\pm i\frac{n}{N}\mbox{ (mod 1)\ \ for $0\le i<N$},
	\label{periodic}
\end{equation}
where $n$ is a positive integer (when $n=0$, it becomes the converged state).
We can call these states as period-$n$ states, and $n$ is a period number.
If `$+$' (`$-$') is used in Eq.\ (\ref{periodic}), $\phi_i$ increases (decreases)
with the same rate as $i$ increases.
Here we look at the stabilities of these period-$n$ states for small $N$'s.
\begin{enumerate}
\item When $N=3$, $E=0$ and $d=1$ (fully connected),
period-1 states [e.g., $(0,\frac{1}{3}, \frac{2}{3})$] are fixed points under both rules;
but they are stable under Rule A, while unstable under Rule B, because, if the difference between
any pair of influencing nodes is greater than or equal to 0.25, where the ${\Delta_B}'(x)$ changes the sign,
the state cannot be stable under Rule B.
\item When $N=4$, $E=0$ and $d=1$, period-1 states [e.g., 
$(0, \frac{1}{4}, \frac{2}{4}, \frac{3}{4})$] are fixed points under both rules; but
they are stable under Rule A, but unstable under Rule B.
\item When $N=4$, $E=2$ and $d=1$ (fully connected), 
period-1 states are unstable fixed points under both rules.
They are unstable under Rule A, because, if the difference between
any pair of influencing nodes is equal to 0.5, where $\Delta_A(x)$ changes the sign when $\theta=0.5$,
the state cannot be stable.
No period-$n$ state ($n>0$) can be stable under Rule B when the network is fully connected,
because there always exist a pair of nodes with the difference greater than 0.25.
\item When $N=5$, $E=0$ and $d=1$, period-1 states are 
stable fixed points under both rules.
\item When $N=5$, $E=5$ and $d=1$ (fully connected), 
period-1 states are fixed points under both rules; but
they are stable under Rule A, while unstable under Rule B.
\item When $N=10$, $E=0$ and $d=1$, both period-1 and
period-2 states are stable fixed points under both rules.
\item When $N=10$, $E=0$ and $d=2$, period-1 states 
are stable fixed point under both rules; but period-2 states are stable 
fixed points under Rule A, while become unstable under Rule B,
because there exist pairs of nodes with the differences at 0.4.
\end{enumerate}
We can easily see that, for fully-connected networks,
converged states are the only stable fixed points under Rule B, while
other stable fixed points can exist under Rule A.
On the other hand, when $E=0$, there exist stable fixed points 
other than converged states under both rules, but 
some stable fixed points under Rule A are unstable under Rule B.
From above examples, it is easy to find that, when $n\ge\frac{N}{2}$ (Rule A) 
or $n\ge\frac{N}{4d}$ (Rule B), period-$n$ states when $E=0$ become unstable.
In the next section, we will find properties of all possible 
stable fixed points in more detail
using numerical simulations, when $N=1000$ and $d\ll N$ with various $e$ values.

\section{Numerical Results}
Once an initial state is given, the state will converge to a fixed point 
eventually. 
Here we will categorize fixed points, and observe what types of 
fixed points are reached for given $e$ under both Rule A and Rule B.
Then, using the Monte-Carlo simulations, we will also estimate ratios of initial 
states, for which a certain types of fixed points will be reached. 
In other words, we will estimate average volumes of the basins of attraction for 
some types of fixed points, which form attractors.
For example, if a network structure is given,
the set of all converged states can be viewed as an attractor for the given dynamical rule, and 
states in a certain portion of the state space will reach converged states 
as time increases.
If we choose $N_r$ random initial conditions from uniformly distributed states and  
$N_0$ of them reach converged states, $N_0/N_r$ is the estimated volume of the basin of attraction for converged states
(the volume of the whole state space, $N$-torus, is 1).
If $N_0/N_r\simeq0$, the volume of the basin of attraction is extremely small, and
the converged state is unlikely to be reached spontaneously in the system
even though the converged states are stable fixed points, unless initial conditions are chosen deliberately.
On the other hand, if $N_0/N_r\simeq1$, the basin of attraction for converged states
can be close to the whole state space.

In our simulations, if $E$ is either 0 or $E_{N,d}$ ($e=0$ or 1), there is only one possibe network configuration.
But if $E$ is between 0 and $E_{N,d}$ ($0<e<1$), there are many possible configurations; hence
we pick both a network configuration for given $E$ and an initial 
condition, randomly out of uniformly distributed choices, for each run.
Then, $N_0/N_r$ will represent estimates of the {\em average} volumes of those basins of attraction
for converged states over all possible network configurations.
The same method can be used on period-$n$ states, too.
We will still use $N_r=1000$ for all our simulations.

\begin{figure}
\begin{center}
\includegraphics[scale=0.55]{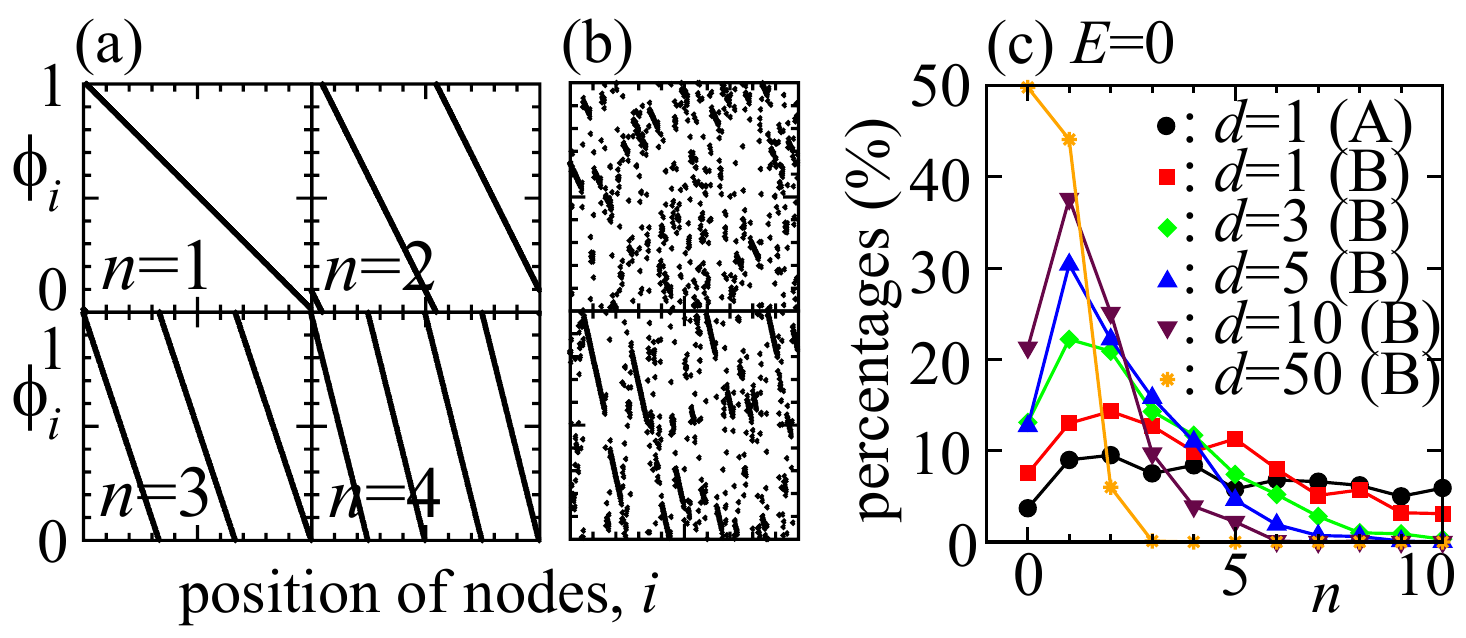}
\caption{\label{E0}
(Color online)
For $N=1000$ and $E=0$, we observe 
(a) examples of period-$n$ states;
(b) examples of typical fixed points when $d>1$ for Rule A; and
(c) ratios of period-$n$ states out of 1000 random initial conditions under
both Rule A ($d=1$) and Rule B (various $d$'s). 
}
\end{center}
\end{figure}

\begin{figure}
\begin{center}
\includegraphics[scale=0.42]{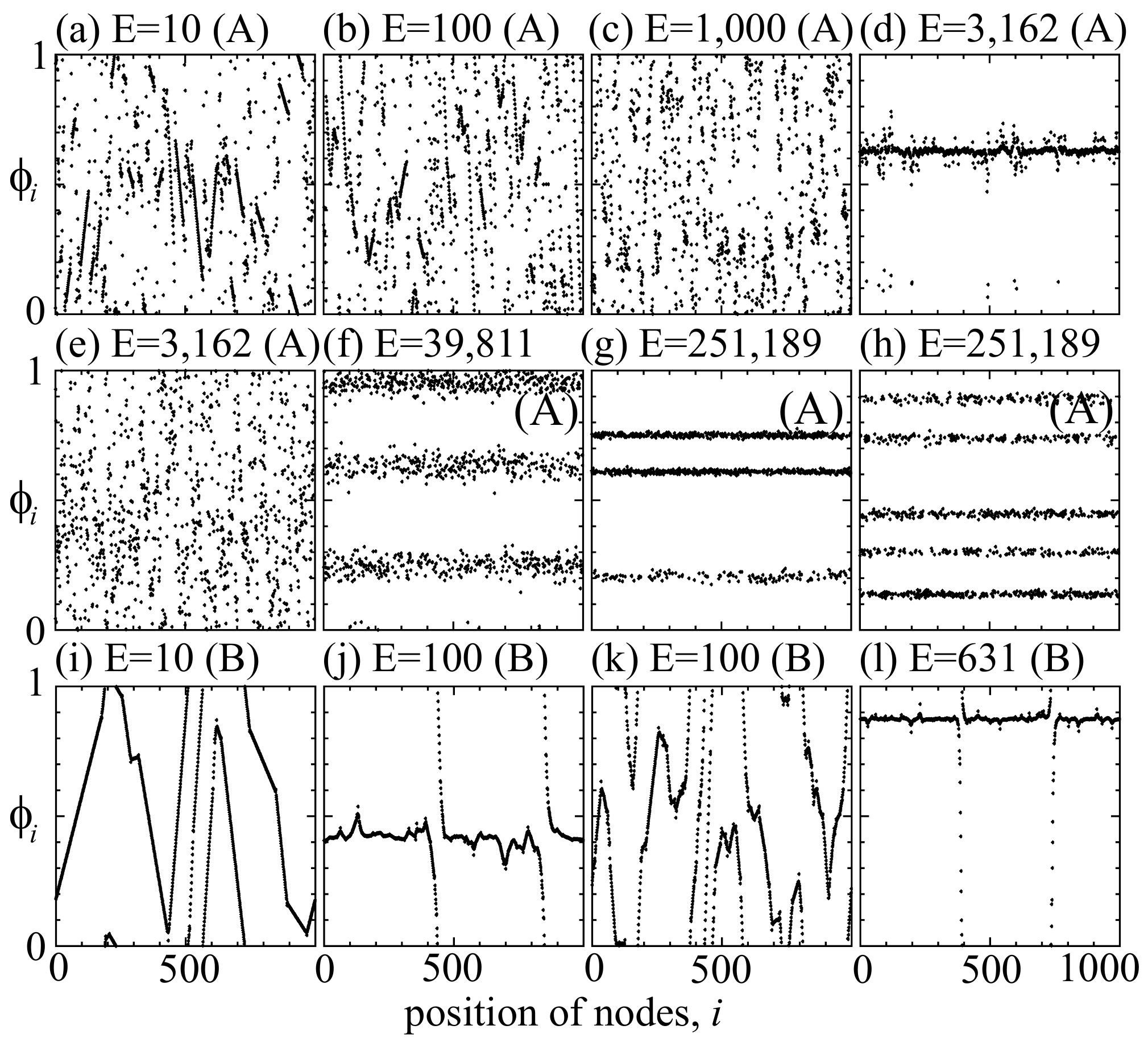}
\caption{\label{states}
For $N=1000$ and $d=3$, we show examples of stable fixed 
points for various $E$ values numerically.
Network configurations and initial conditions are randomly picked for given $E$'s.
(a)-(h) Under Rule A, 
(i)-(l) Under Rule B.
}
\end{center}
\end{figure}

First, we find fixed points when $N=1000$ and $E=0$ (see Fig.\ \ref{E0}).
Typical fixed points here are period-$n$ states
defined in Eq.\ (\ref{periodic}).
For $d=1$ under Rule A and any $d$ under Rule B, 
period-$n$ states are only fixed points found [see Fig.\ \ref{E0}(a)],
but for $d>1$ under Rule A, other types of fixed points exist as shown 
in Fig.\ \ref{E0}(b) [meaning that 
Eq.\ (\ref{periodic}) is not the only solution that satisfy 
$\forall i, \phi_i(t+1)=\phi_i(t)$].
Actually they dominate because no period-$n$
state was reached from 1000 randomly given initial conditions
when $k>1$ under Rule A.
These states have small {\em locally-converged}\footnote{
It means differences between states of neighbors (for example, 
nodes $i-1$ and $i+1$ for node $i$) are very small.
If a state is locally converged throughout the whole structure, 
the state represented in the scatter plots of $\phi_i$ versus $i$ in Fig.\ \ref{states} will
look like a one-dimensional curve.}
segments with scattered values
In Fig.\ \ref{E0}(c), we observe percentages of period-$n$ states, including
converged states ($n=0$), out of 1000 random initial conditions.
The higher $n$, the less likely they will be reached from random
initial conditions in most cases.
Under Rule B, as $d$ increases, the highest $n$ that can be found in fixed points
decreases, and the volume of the basin of attraction for converged states increases;
for example, when $d=50$, about 50\% of initial conditions reach converged states.
For $d=2$ and 3 under Rule A, percentages of period-$n$ states including 
converged states are all zero in our simulations.

Next, we observe examples of fixed points under both rules as we vary $e$ from 0 to 1
for the case of $N=1000$ and $d=3$
in Fig.\ \ref{states}.
If we look at fixed points under Rule A, all fixed points are similar to
states in Fig.\ \ref{E0}(b) 
when $E$ is less than about $1000$
as shown in Figs.\ \ref{states}(a)-\ref{states}(c).
As $E$ increases further, gathered states like the one shown in Fig.\ \ref{states}(d) 
start to appear, while diversely spread states still exist as in Fig.\ \ref{states}(e) 
[both are at $E=3162$ ($e\simeq6.5\times10^{-3}$)].
Converged states starts to appear when $E\simeq4000$, and starts to dominate 
at $E\simeq20\ 000$ with about 97\% of all fixed points.
At this point, another types of fixed points called {\em grouping}
states start to appear.
In these states, nodes are divided into separated groups of nodes with
clustered state values, and they can be called as 
$m$-group states, where $m$ stands for the number of groups.
For the case of  $\theta=0.5$ (Rule A),  
even-numbered groups are not reached\footnote{
For even-numbered groups, there exist
two groups that have the difference of states at 0.5, 
and that is why these states are mostly unstable.},
while, under Rule B, grouping states are unstable, and will not be reached.
When $m=1$, states are just converged states, so 3-group states
are the ones that appear mostly [see Figs.\ \ref{states}(f) and \ref{states}(g)].
Widths of state values for groups tend to decrease as $E$ increases further,
reaching 0 at $E=E_{N,d}$,
while grouping states with $m\ge5$ also starts to appear at $E\simeq120\ 000$
[see Fig.\ \ref{states}(h)].

Under Rule B, fixed points look a little different.
First of all, stable fixed points are mostly locally converged throughout the 
whole structure, showing piecewise-linear shapes in the 
scatter plots in Figs.\ \ref{states}(i), \ref{states}(j), and \ref{states}(k) 
when $E=10$ and 100.
In these states, some can be gathered 
as in Fig.\ \ref{states}(j) and 
others can be spread as in Fig.\ \ref{states}(k), even
at the same $E$ value.
Under Rule B, converged states starts to appear at $E\simeq500$,
and reach 100\% at $E\simeq1200$; while gathered states 
as in Fig.\ \ref{states}(l) at $E=631$ and 
converged states coexist in this region.
As $E$ further increases, all states will lead to converged states.
 
If we categorize fixed points based on our numerical experiments
so far, they are as below.
\begin{enumerate}
\item Converged states: 
All nodes have the same state value in these states, and they are 
stable in all cases even though the volume of the basin of
attraction for these states can be almost zero when $e\ll1$. 
All converged states satisfy 
\begin{equation}
	\phi_0=\phi_1=\phi_2=\cdots=\phi_{N-1}.
	\label{converged_states}
\end{equation}
A set of all possible converged states can be viewed as an attractor,
and the volume of the basin of attraction will be estimated numerically.
\item Periodic states: In these states, $\phi_i$'s increase (or decrease) 
linearly as $i$ increases along the structure with $n$ cycles, 
and they are called as period-$n$ states, where a positive integer
$n$ is in the range of $0<n<\frac{N}{2}$ (Rule A) 
or $0<n<\frac{N}{4k}$ (Rule B).
We exclude $n=0$, because period-0 states are converged state. 
These states are stable fixed points when $E=0$ (when there is no long-range
link). An equation for the period-$n$ states is given in 
Eq.\ (\ref{periodic}).
For each $n$, there are two attractors, one with the set of increasing
period-$n$ states, and the other with the set of decreasing period-$n$ states.
Here we will estimate the sum of volumes of basins of attraction for these two attractors
for each $n$.
\item Grouping states: Nodes can be divided into separated 
groups of nodes with
clustered state values, and they can be called as 
$m$-group states, where $m$ stands for the number of groups.
We exclude $m=1$, because 1-group states are converged states.
Grouping states can be found only when $e\simeq1$ under Rule A.
If $e=1$, where the network is fully connected,
each group will consist of nodes with the same value.
Groups don't have to be equally distanced; rather
it depends on sizes of groups as shown in Fig.\ \ref{states}(g).
\item Other states: 
There can be fixed points that don't belong to above categories, and
nodes look mostly scattered in the scatter plots.
We can find some fixed points when $N$ is small.
For example, in Fig.\ \ref{model}(b), cases with 
$N=4$ and 5 are shown, and
we can find some fixed points under Rule A.
For the case of $N=4$, $E=1$ and $k=1$, 
$(0,\frac{1}{8},\frac{1}{2},\frac{7}{8})$ is a fixed point; 
for the case of $N=5$ $E=1$ and $k=1$, 
$(0,\frac{1}{11},\frac{4}{11},\frac{7}{11},\frac{10}{11})$ is a fixed point; 
and for the case of $N=5$, $E=2$ 
and $k=1$, $(0,\frac{1}{3},\frac{1}{2},\frac{1}{2},\frac{2}{3})$ is a fixed point. 
For $N\gg1$ and  $E\gg1$, these fixed points can be highly complicated 
(see Fig.\ \ref{states}).
If we categorize these states further, there can be two different types.
\begin{itemize}
	\item Gathered states: As shown in Figs.\ \ref{states}(d) 
		and \ref{states}(j),
		state variables are gathered near one value, even though
		some states can be far from this value.
	\item Spread states: As shown in Figs.\ \ref{states}(e) and \ref{states}(k),
		state variables are spread well between 0 and 1, even though 
		we can find some patterns in the scatter plots.
\end{itemize}
\end{enumerate}

We can elaborate more on two types of {\em other} states mentioned above.
Under Rule A, when $E$ is less than about 5000,
the whole spectrum of states between these two types of states can
be reached.
But for the approximate range of 
$5\times10^3<E<1.5\times10^4$ ($10^{-2}<e<3\times10^{-2}$), they are clearly separated (see Fig.\ 6).
Under Rule B, though, states are locally converged throughout the whole 
structure; therefore 
states are represented by an one-dimensional curves in scatter plots. 
For low $E$'s, these curves are piecewise linear, and
points where linear segments meet are the locations where
long-range connections branch off.
If $E$ is greater, the numbers of these branching points 
increase, too, and the curves become more complicated.
As in Rule A, gathered and spread states can coexist for the cases with the same $E$.

To see how nodes are spread for these fixed points,
we introduce a quantity $s$.
First, we need an anchor point because we learned in the previous section
that an anchor point is necessary to find the average value of $\phi_i$'s.
Here we set the anchor point $\phi_s$ as 
the point where the most nodes are gathered state-wise;
in other words, if we find the probability distribution function, $P(\phi)$,
the anchor point $\phi_s$ is the point where $P(\phi_s)$ is the maximum
for a given fixed point.
Second, we can find the average of all $\phi_i$'s, $\langle\phi\rangle_s$, viewed from $\phi_s$,
using $\Delta_A(x)$ with $\theta=0.5$,
\begin{equation}
	\langle\phi\rangle_s=\phi_s+\frac{1}{N}\sum_{j=0}^{N-1}
	\Delta_A(\phi_j-\phi_s)\ \mbox{(mod 1)}.
	\label{average_s}
\end{equation}
Finally, $s$ is defined as 
\begin{equation}
	s\equiv\sqrt{\frac{12}{N}\sum_{j=0}^{N-1}
	[\Delta_A(\phi_j-\langle\phi\rangle_s)]^2}.
	\label{s}
\end{equation}
This is the standard deviation of node states viewed from $\phi_s$\footnote{
To make $s$ for uniformly distributed states normalized, the standard deviation has
been multiplied by $\sqrt{12}$.}.
For converged states, $s$ should be 0, while, for uniformly distributed
states like period-$n$ states, $s$ is 1.
For other states, if it is close to 0, the state is gathered near one value;
and if it is close to 1, the state is diversely spread.
(Note that 1 is not the maximum possible value.)
Using this index, we can see distributions of various fixed points in more detail.

\begin{figure}
\begin{center}
\includegraphics[scale=0.42]{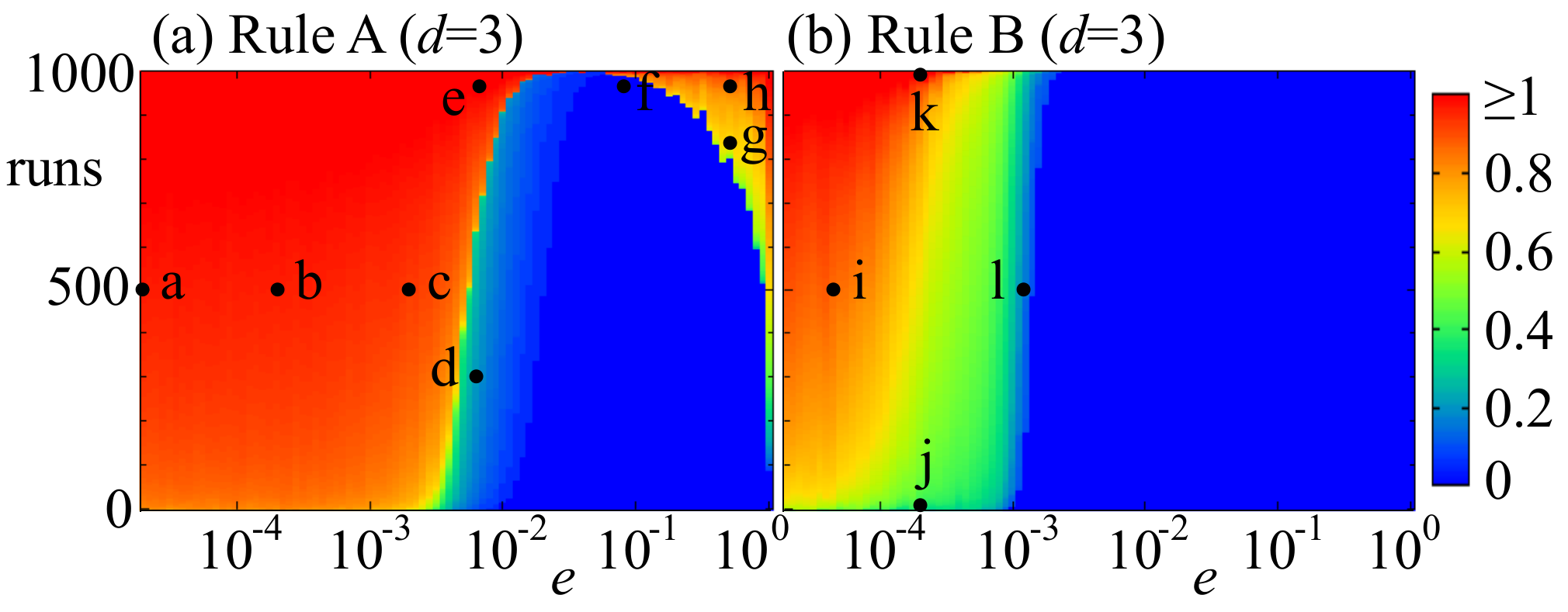}
\caption{\label{diversity}
(Color online)
We observe $s$ for 1000 runs 
for various $E$ values ($E\ge10$) when $N=1000$ and $d=3$. 
For given $E$, both positions of long-range links and an initial condition were
chosen randomly for each run.
Each pixel represents a run, and a column of pixels, representing runs for given $E$,
are ordered, so that the run with the lowest $s$ is in the bottom.
Blue regions represent converged states ($s=0$), and red regions represent
quite diverse states ($s\ge1$). 
(a) Values of $s$ under Rule A.
(b) Values of $s$ under Rule B.
There are 12 dots in graphs, indexed by alphabets, and they correspond 
to the states depicted in Fig.\ \ref{states}.}
\end{center}
\end{figure}

In Fig.\ \ref{diversity},
we observe values of $s$ using Monte-Carlo simulations, which show us
what types of fixed points are more likely to appear spontaneously.
Under Rule A [see Fig.\ \ref{diversity}(a)], 
all fixed points are spread states
when $E$ is less than about $1000$ ($e\simeq2\times10^{-3}$).
But a transition occurs in the approximate range of  $10^3<E<2\times10^4$
($2\times10^{-3}<e<4\times10^{-2}$),
and gathered states start to appear.
At $E\simeq4\times10^3$ ($e\simeq8\times10^{-3}$), 
converged states start to appear, too.
All three kinds of fixed points coexist until $E\simeq2\times10^4$ ($e\simeq
4\times10^{-2}$), where converged states are dominant (about 97\%) and
spread states no longer exist.
At this point, grouping states start to appear, and 3-group states
are the ones that appear mostly.
For $2\times10^4<E<5\times10^4$ ($4\times10^{-2}<e<10^{-1}$), converged states 
dominate and reach about 99\%.
Gathered and spread states have all disappeared,
and grouping states start to increase as $E$ further increases.
Grouping states with $m\ge5$ also starts to appear at $E\simeq1.2\times10^5$
($e\simeq2.4\times10^{-1}$).
Grouping states are  in the top-right corner
of Fig.\ \ref{diversity}(a).

Under Rule B, gathered and spread states also appear when $E$ is small,
even though states are mostly one-dimensional in scatter plots as seen in Fig.\ \ref{states}.
Converged states start to appear
at $E\simeq500$ 
($e\simeq10^{-3}$), and their distributions reach 100\% at $E\simeq1200$ ($e\simeq
2.4\times10^{-3}$).
Note that more than 10\% of initial conditions reach converged states 
when $E=0$ as shown in Fig.\ \ref{E0}(c).
But those converged states disappear before $E$ reaches 10.
When $E$ is greater than 1200, only converged states are observed, meaning that  
the volume of the basin of attraction for converged states is the volume of
the whole state space.
There are no grouping states here as discussed earlier.
One interesting point is that, for the region of $e$ for small-world behavior (when $e$ is less than about $10^{-3}$),
reached fixed points are all diverse states without any converged state.

Now we turn our attention to the case of $d=0$, where all links are randomly attached.
It is easy to see that a necessary condition for a network to have 
fixed points other than converged states is for the network to be cyclic.
For acyclic networks, only converged states are fixed points. 
In our model for $d>0$, short-range links guarantee that these networks
have at least a cycle, and we observed many types of fixed points so far.
What happens if there is no prearranged short-range links for the case of $d=0$?
First, there can be more than one component, a group of connected nodes, for low $E$'s;
hence we only consider states of nodes for the largest component when calculating $s$ and
$P(\phi)$, the probability distribution function of $\phi_i$'s for a given fixed point, since we are interested in looking at
how connected nodes influence each other.
Second, the chance of the largest component of the random network to be acyclic is not negligible
when $E$ is low, but, as $E$ increases, that chance becomes extremely small.
But, for higher $E$'s, we expect to see similar results as those for the cases of $d>0$,
because influences of short-range links become negligible when $E\gg dN$.

\begin{figure}
\begin{center}
\includegraphics[scale=0.50]{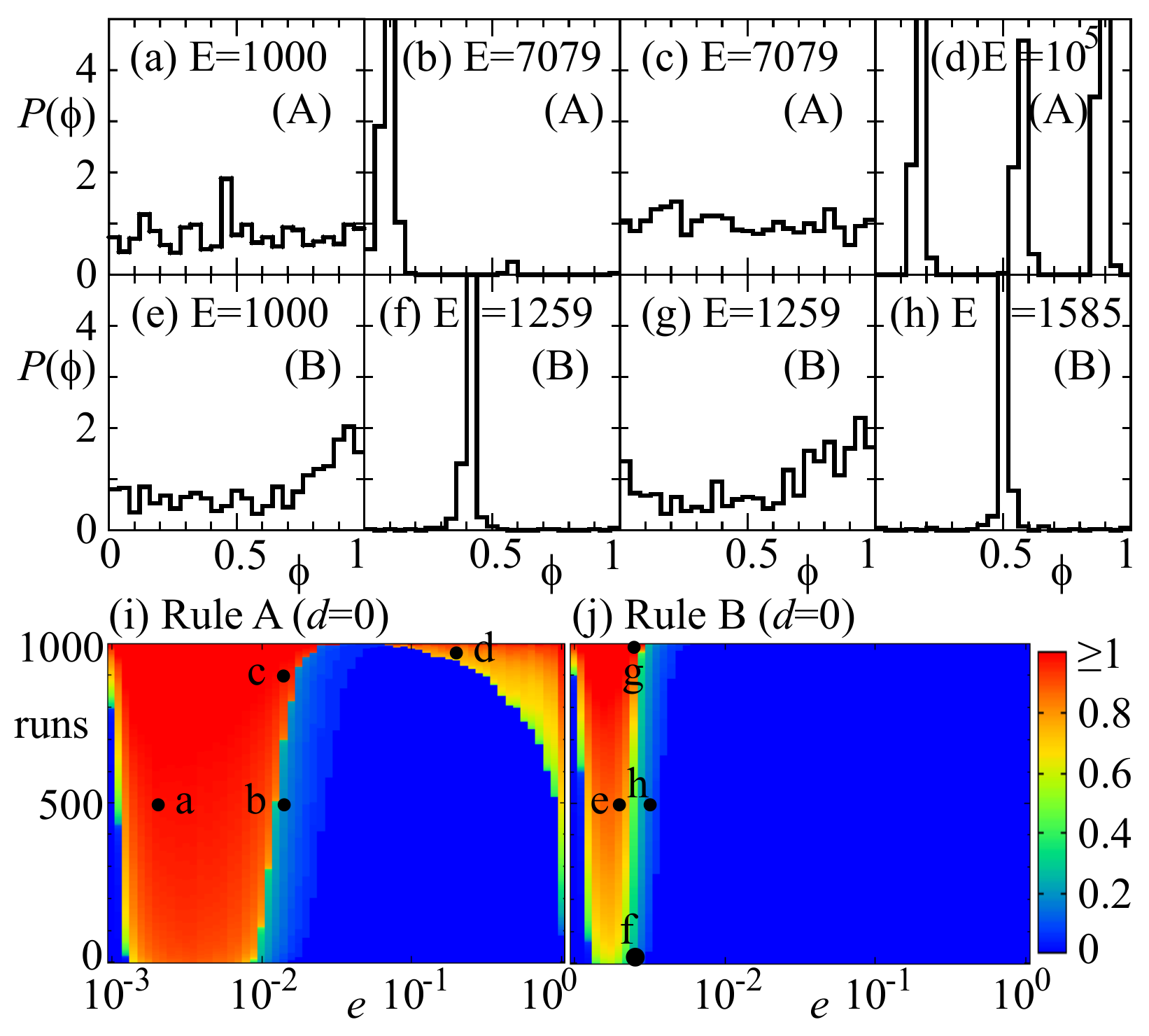}
\caption{\label{random}
(Color online)
We observe $s$ for 1000 runs 
for various $E$ values ($E\ge500$) when $N=1000$ and $d=0$. 
(a)-(d) Distributions of states $P(\phi)$ under Rule A.
(e)-(h) Distributions of states $P(\phi)$ under Rule B.
(i) Values of $s$ under Rule A.
(j) Values of $s$ under Rule B.
There are 8 dots in graphs, indexed by alphabets, and they correspond 
to the states in this figure.}
\end{center}
\end{figure}

In Fig.\ \ref{random},
we look at some examples of fixed points for the case of $N=1000$ and $d=0$, 
and observe behavior of $s$ as in Figs.\ \ref{states} and \ref{diversity}. 
Since scatter plots of $\phi_i$ versus $i$ are no longer meaningful, we use the probability
distribution function $P(\phi)$ to represent each fixed point. 
In Figs.\ \ref{random}(a)-\ref{random}(h), some fixed points
under both rules are shown.
We can find all types of fixed points except the periodic states.
We can also observe 2-group states under Rule A here, unlike the cases of $d>0$ [for 2-group states,
a group usually consists of only few nodes as shown in Fig.\ \ref{random}(b)].
In Figs.\ \ref{random}(i) and \ref{random}(j), we observe $s$ values for 1000 runs for each $E$ under both rules.
When $E$ is less than about 600 ($e<1.2\times10^{-3}$), converged states are dominant as stable
fixed points, but,
as $E$ increases further, diverse states appear and later disappear with the same kind of transition
from diverse states to converged states as we have already observed for cases of $d=3$ in Fig.\ \ref{diversity}.

\begin{figure}
\begin{center}
\includegraphics[scale=0.5]{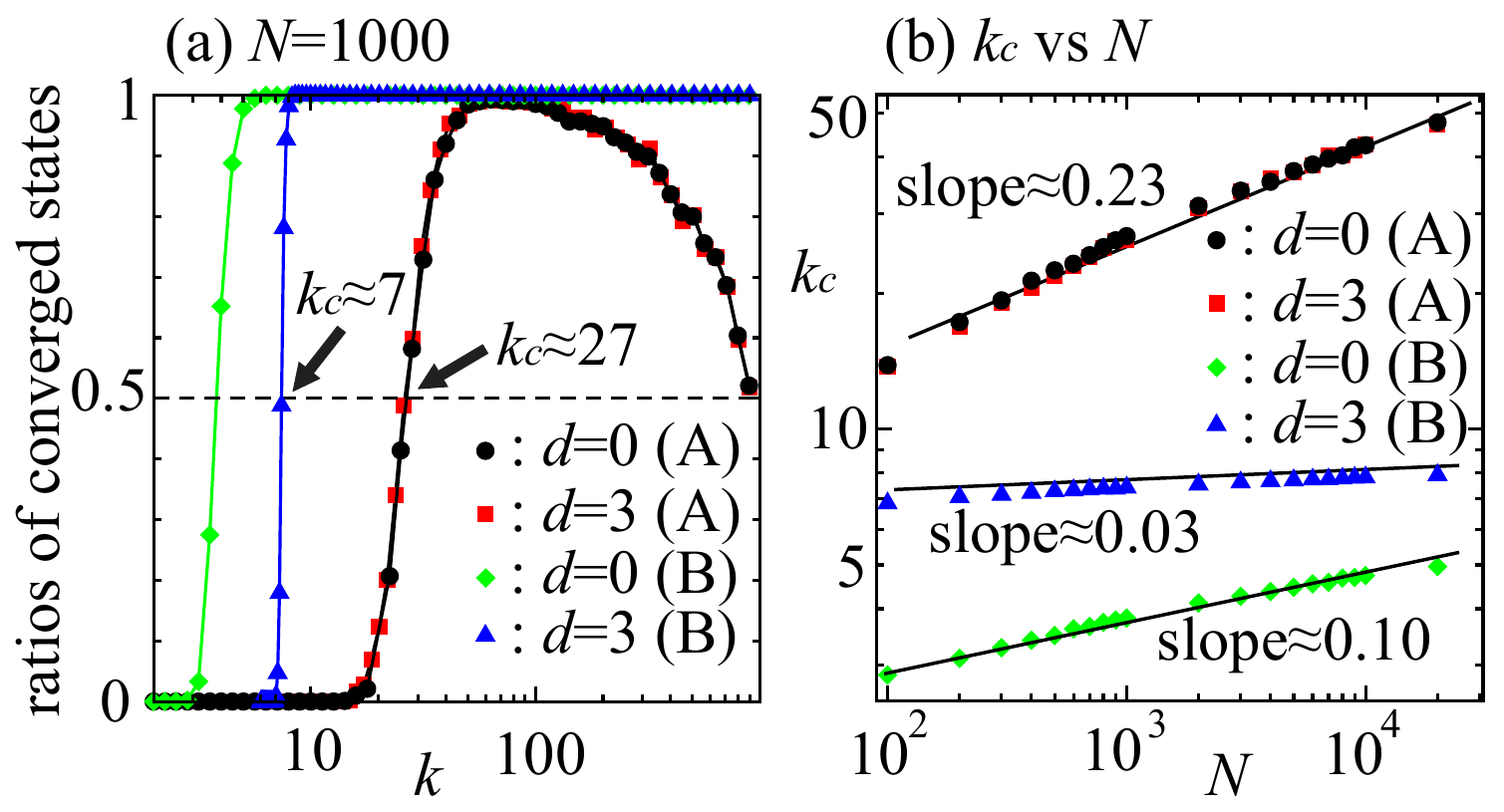}
\caption{\label{scale}
(Color online)
We observe values of $E$ at which converged states start to prevail under both rules.
(a) Ratios of converged states out of 1000 runs versus $k$, the average connectivity,
when $d=0$ and 3 under both rules. Values of $k$ at which ratios reach 0.5 will be called as $k_c$.   
(b) $k_c$ versus $N$ for four cases. Slopes are exponents, $\xi$'s, in $k_c\propto N^\xi$.
}
\end{center}
\end{figure}

One important dynamic property on networks we have been using is this transition from 
diverse states to converged states as the number of links $E$ increases. 
(Under Rule A, grouping states appear after this transition.) 
Then, we can ask when this transition occurs.
One way to quantify the answer is finding the point at which the ratio of converged states out of all runs
becomes exactly a half, and we will use the average connectivity, instead of $E$, to represent
this value: $k_c$ [note that $k=2(d+E/N)$].
In Fig.\ \ref{scale}(a), 
we show ratios of converged states out of 1000 runs versus $k$ for four cases we have used at $N=1000$: 
$d=0$, 3 (Rule A), and $d=0$, 3 (Rule B).
Under Rule A, two curves match very well, and $k_c$ is about 27;
while under Rule B, two curves don't match because the numbers of nodes and links for the largest
component are much smaller than $N$ and $E$ at $k\simeq k_c$ when $d=0$.
(For the case of $d=0$ under Rule A, on the other hand, 
the possibility of having more than one component is extremely low at $k\simeq k_c$.)
Finally, in Fig.\ \ref{scale}(b), we observe how $k_c$ changes as $N$ increases.
Overall dynamic behavior will not be much different, but how $k_c$ varies with $N$?
Numerical results show that $k_c$ fit well with $N^\xi$, and exponents $\xi$'s for four cases
are also found.
Exponents don't seem to depend on $d$ under Rule A, and $\xi\simeq0.23$; while
under Rule B, $\xi\simeq0.10$ ($d=0$) and $\xi\simeq0.03$ ($d=3$).
Results suggest that more long-range links are needed to get converged states under Rule A,
and $k_c$ increases faster with $N$, compared to results under Rule B.
It is also interesting that, under Rule B ($d=3$), $k_c$ doesn't increase much with $N$ (for example, 
$k_c\simeq7.4$ at $N=10^3$ and $k_c\simeq7.8$ at $N=10^4$).

\section{Discussions}
Using a simple network model, we have found fixed points
for periodic state variables using two dynamic rules of repeated averaging, Rule A and Rule B.
This model has a parameter $e$ ($0\le e \le1$), representing
the number of long-range links, and stable fixed points and their likelihood 
of appearing spontaneously from random initial conditions have changed as $e$ varies.
We categorized fixed points as converged states, periodic states, grouping states, and 
other states (gathered and spread).
Grouping states are only reached under Rule A, 
while other types of fixed points have been found under both rules.
As the number of long-range links increases, the transition from diverse states to converged states 
occurred in all cases,
and we represented the point of transition using $k_c$, the average connectivity 
when the ratio of converged states becomes exactly 0.5. 
Numerically found $k_c$'s fit well with $N^\xi$, and exponents, $\xi$'s, 
have been found.
For Rule A, we need more long-range connections (higher $k_c$) to get converged states,
and the exponent $\xi$ is greater, compared to results under Rule B.
In addition, diverse states can reappear in the form of grouping states under Rule A, 
when the network is close to being fully connected.

The same dynamic rule (Rule A) has been used in previous works\cite{Ree:2011,Ree:2011a}
for dynamic interaction networks with random-walking nodes in two-dimensional lattices.
In these networks, short-range links constantly change as nodes move randomly
in the lattice, thereby giving stochasticity to the model.
Unlike the static network used here, converged states are the only stable fixed points, and 
periodic and grouping states are found to be metastable with finite lifetimes.
For dynamic cases, periodic states are metastable when $0<e\ll1$,
while they can be fixed points when $e=0$ here.
In addition, fixed points like gathered and spread states cannot be stable in dynamic networks,
because these states are only stable specifically for given network configurations.

The motivation for this work is to answer the questions of how the diversities exist in reality even though 
many mechanisms of assimilation are at work, 
and how these diversities disappear if we increase long-range interactions.
For example, if we look at human cultures, they are still quite diverse, but neighboring cultures 
differ only slightly because neighbors usually interact and assimilate.
One can use interaction networks to describe these systems, and argue that 
the prevalence of short-range interactions are the main cause of these kinds of diversities.
Our model showed that, if there are only short-range links, periodic states, which are locally
converged and yet diverse, can be stable fixed points. 
The point of transition from diversity to convergence we found in this model can be used as 
a guide when we roughly determine how many long-range links are necessary to make a group of
nodes converge to one value. 
Even though the reality is more complex, we hope this simple model can 
give us insight when we try to understand phenomena of sustained diversities and their breakdown.

\section*{Acknowledgments}
This work was supported by Kongju National University.

\bibliography{p14}

\end{document}